\documentclass[12pt,a4paper]{article}
\usepackage{verbatim} % comment-environment
\usepackage{amsmath} %
\usepackage{amssymb} %
\usepackage{theorem} %
\usepackage[latin1]{inputenc}
\newcommand{\di}{\,\mathrm{d}}

\newcommand{\R}{\mathbb{R}}

\theorembodyfont{\itshape} \theoremstyle{plain}
\newtheorem{Theorem}{Theorem}[section]
\newtheorem{Lemma}{Lemma}[section]

\author{Helmut Linde} %
\title{A lower bound for the ground state energy of a Schr\"odinger operator on a loop.}
\date{\small\itshape
Department of Physics, Pontificia Universidad Cat\'olica de Chile Casilla 306, Correo 22 Santiago, Chile.}

\begin{document}
\maketitle

\begin{abstract}
\noindent Consider a one dimensional quantum mechanical particle described by the Schrödinger equation on a closed
curve of length $2\pi$. Assume that the potential is given by the square of the curve's curvature. We show that in this
case the energy of the particle can not be lower than $0.6085$. We also prove that it is not lower than $1$ (the
conjectured optimal lower bound) for a certain class of closed curves that have an additional geometrical property.
\end{abstract}

%%%%%%%%%%%%%%%%%%%%%%%%%%%%%%%%%%%%%%%%%%%%%%%%%%%%%%%%%%%%%%%%%%%%%%%%%%%%%%%%%%%%%%%%%%%%%%%%%%%%%%%%%%%%%%%%%%%%%%%%
\section{Introduction}
%%%%%%%%%%%%%%%%%%%%%%%%%%%%%%%%%%%%%%%%%%%%%%%%%%%%%%%%%%%%%%%%%%%%%%%%%%%%%%%%%%%%%%%%%%%%%%%%%%%%%%%%%%%%%%%%%%%%%%%%

Let $\Gamma$ be a smooth closed curve of length $2\pi$ in the plane with the curvature $\kappa(s)$ which is regarded as
a function of the arc length. We consider the Schr\"odinger operator
$$H_\Gamma = - \Delta  + \kappa^2(s) \quad \mathrm{in} \quad L^2([0,2\pi)) $$
with periodic boundary conditions. Let $\lambda_\Gamma$ be the lowest eigenvalue of $H_\Gamma$. It has been conjectured
that $\lambda_\Gamma \ge 1$ for any $\Gamma$. The class ${\cal F}$ of the conjectured minimizers of $\lambda_\Gamma$
contains the circle and certain point-symmetric oval loops. For all curves in ${\cal F}$ the equality $\lambda_\Gamma =
1$ holds, but so far it has not been shown that this is actually the smallest possible value of $\lambda_\Gamma$. In
their paper \cite{BL} Benguria and Loss established a connection between this problem and the Lieb-Thirring conjecture
in one dimension. They also proved that $\lambda_\Gamma \ge 0.5$, which seems to be the best lower bound for $\lambda_\Gamma$
so far.\\
Recently, Burchard and Thomas have shown \cite{BT} that the curves in ${\cal F}$ minimize $\lambda_\Gamma$ at
least locally, i.e., there is no small variation around these curves that reduces $\lambda_\Gamma$.\\
In the present article we will add further credibility to the mentioned conjecture in two ways. On the one hand, we
show that $\lambda_\Gamma \ge 1$ holds for a considerable class of curves that meet a certain additional geometrical
condition. Extending this method to the class of all curves of interest yields, on the other hand, an improved lower
bound on $\lambda_\Gamma$.
%%%%%%%%%%%%%%%%%%%%%%%%%%%%%%%%%%%%%%%%%%%%%%%%%%%%%%%%%%%%%%%%%%%%%%%%%%%%%%%%%%%%%%%%%%%%%%%%%%%%%%%%%%%%%%%%%%%%%%%%
\section{Statement of the result}
%%%%%%%%%%%%%%%%%%%%%%%%%%%%%%%%%%%%%%%%%%%%%%%%%%%%%%%%%%%%%%%%%%%%%%%%%%%%%%%%%%%%%%%%%%%%%%%%%%%%%%%%%%%%%%%%%%%%%%%%

For a given smooth curve $\Gamma$ with an arc length parameter $s$ we introduce the angle $\phi(s)$ between the tangent
on $\Gamma$ in $s$ and some fixed axis, which implies $\phi'(s) = \kappa(s)$. For the sake of simplicity we will only
consider strictly convex curves, i.e., $\phi' > 0$. To keep the notation compact we write
$$\phi: \Omega \rightarrow \Omega \quad \mathrm{with}\quad\Omega := \R \slash 2\pi \mathbb Z,$$
considering numbers that differ by an integer multiple of $2\pi$ as identical. Our main result is:
\begin{Theorem} \label{Theorem1}
Let $\Gamma$ be a smooth, strictly convex, closed curve of length $2\pi$ in the plane and $\lambda_\Gamma$ defined as
above. Then $$\lambda_\Gamma > \left(1+\frac{1}{1+8/\pi}\right)^{-2} > 0.6085.$$
\end{Theorem}
In the proof of Theorem \ref{Theorem1} we will make use of the following geometrical concept: We call $s \in \Omega$ a
`critical point' of $\Gamma$ if $\phi(s + \pi) = \phi(s) + \pi$. Obviously, $s+\pi$ also is a critical point then. If
$s$ is a critical point, we call $\phi(s)$ a `critical angle'. While open curves may have no critical points at all,
the following lemma holds for the closed curves we are considering:
\begin{Lemma} \label{LemmacriticalPoints}
Every smooth closed curve $\Gamma$ has at least six critical points.
\end{Lemma}
It is clear from the definition of a critical point and the lemma, that every $\Gamma$ has at least three critical
points and three critical angles in $[s,s+\pi) \subset \Omega$ for any $s \in \Omega$. For a class of curves that have
their critical angles distributed somewhat evenly we can show that $\lambda_\Gamma \ge 1$ holds:
\begin{Theorem} \label{Theorem2}
Let $\Gamma$ be as in Theorem \ref{Theorem1} and assume additionally that every interval $[\phi,\phi+\frac\pi 2)
\subset \Omega$ contains at least one critical angle of $\Gamma$. Then $\lambda_\Gamma \ge 1$.
\end{Theorem}
It is an immediate consequence of Theorem \ref{Theorem2} and Lemma \ref{LemmacriticalPoints}, that for any hypothetical
curve $\Gamma$ with $\lambda_\Gamma < 1$ there is a $\phi$ such that $[\phi,\phi+\frac\pi 2)$ and
$[\phi+\pi,\phi+\frac{3\pi}2)$ each
contain at least three critical angles and $[\phi+\frac\pi 2,s+\pi) \cup [\phi+\frac{3\pi}{2},\phi+2\pi)$ none.\\
A few comments on the geometrical interpretation of the above said are in order: Although we have defined the critical
points for a certain parameterization of the curve, the location of the critical points is an intrinsic property of the
curve. More precisely, two points $P_1$ and $P_2$ on a closed curve of length $2\pi$ are critical, if the arc length
between $P_1$ and $P_2$ is $\pi$ and the tangents in these two points are parallel.\\
From Theorem \ref{Theorem2} follows that only curves with a rather uneven distribution of their critical points are
candidates for $\lambda_\Gamma < 1$.  Roughly speaking, such curves tend to be rather symmetric, as will become clear
in the proof of Theorem \ref{Theorem1}. In fact, the function $f$, that we will define and estimate in the proof, can
in some sense be seen as a measure for how far $\Gamma$ is away from being point-symmetric. This, on the other
hand, will enable us to estimate how far $\lambda_\Gamma$ could be below one.\\
We are not aware of any direct correlation between the distribution of the critical points and the ground state energy
$\lambda_\Gamma$, except the connection that is established by Theorem \ref{Theorem2}, of course. There seems to be no
reason why the condition of Theorem \ref{Theorem2} should `prefer' curves with a high energy, especially if one takes
into account that the conjectured minimizers meet this condition. We believe that this makes the conjecture
$\lambda_\Gamma \ge 1$ even more credible.\\
The remainder of the article is devoted to proving Lemma \ref{LemmacriticalPoints} and the two theorems.

%%%%%%%%%%%%%%%%%%%%%%%%%%%%%%%%%%%%%%%%%%%%%%%%%%%%%%%%%%%%%%%%%%%%%%%%%%%%%%%%%%%%%%%%%%%%%%%%%%%%%%%%%%%%%%%%%%%%%%%%
\section{Proof of the results}
%%%%%%%%%%%%%%%%%%%%%%%%%%%%%%%%%%%%%%%%%%%%%%%%%%%%%%%%%%%%%%%%%%%%%%%%%%%%%%%%%%%%%%%%%%%%%%%%%%%%%%%%%%%%%%%%%%%%%%%%

To prepare the proofs of Lemma \ref{LemmacriticalPoints} and the two theorems we introduce some more notation: We
consider a curve $\Gamma$ as in Theorem $\ref{Theorem1}$ and assume without loss of generality that $\phi' > 0$.
Because $\Gamma$ is closed $\phi$ meets the conditions
\begin{eqnarray*}
&&\int_\Omega \cos \phi(s) \di s =  \int_\Omega \sin \phi(s) \di s = 0,\\
&&\int_\Omega \phi'(s) \di s = 2\pi.
\end{eqnarray*}
We note that $\phi(s)$ has an inverse function $\phi^{-1}: \Omega\rightarrow \Omega$, and the closure conditions are
equivalent to
\begin{eqnarray*}
&&\int_\Omega (\phi^{-1})'(t) \sin t \di t =  \int_\Omega (\phi^{-1})'(t) \cos t \di t = 0,\\
&&\int_\Omega (\phi^{-1})'(t) \di t = 2\pi.
\end{eqnarray*}
The function $(\phi^{-1})'$ can therefore be written as a Fourier series
$$(\phi^{-1})'(t) = 1 + \sum\limits_{n=2}^\infty n a_n \cos nt - n b_n \sin nt,$$
such that
$$\phi^{-1}(t) = C + t + \sum\limits_{n=2}^\infty a_n \sin nt + b_n \cos nt.$$
By the invariance of the problem under a shift of the arc length parameter $s$ we can assume that $C=0$. Then we can
write $\phi^{-1}$ in the form
\begin{equation}\label{EqPhi}
\phi^{-1}(t) = t + g(t) + f(t),
\end{equation}
where
\begin{eqnarray*}
g(t) &:=& \sum\limits_{n=2,4,6,...}^\infty a_{n} \sin nt + b_{n} \cos nt,\\
f(t) &:=& \sum\limits_{n=3,5,7,...}^\infty a_{n} \sin nt + b_{n} \cos nt.
\end{eqnarray*}
Note that
\begin{equation}\label{EqImparityOfF}
f(t+\pi) = - f(t),\quad g(t+\pi) = g(t) \quad \textmd{for all}\quad t\in\Omega.
\end{equation}
Proof of Lemma \ref{LemmacriticalPoints}: From (\ref{EqPhi}) and (\ref{EqImparityOfF}) it is easy to see that the
critical angles of $\Gamma$ are just the zeroes of $f$. By continuity of $f$ and (\ref{EqImparityOfF}), any nontrivial
$f$ clearly has at least two zeroes $t_0$ and $t_0+\pi$ in $\Omega$ with a change of sign. But if these were the only
zeroes, we would have
$$\int_\Omega f(t) \sin (t-t_0) \di t \neq 0,$$ which is impossible by the definition of $f$. So $f$ must change its sign in at least
one more point. By the symmetry property (\ref{EqImparityOfF}) it is clear that if, say, $f(t_0+\epsilon) > 0$ then
$f(t_0+\pi-\epsilon) > 0$ for small $\epsilon > 0$. That means that each of the intervals $(t_0,t_0+\pi)$ and
$(t_0+\pi,t_0+2\pi)$ contains an even number of zeroes with a change of sign. In total, this leads to a minimum of six
zeroes of $f$ with a change of sign.
\begin{flushright} $\blacksquare$ \end{flushright}
We now state and prove a lemma that is key to the proofs of Theorem \ref{Theorem1} and Theorem \ref{Theorem2}.
%%%%%%%%%%%%%%%%%%%%%%%%%%%%%%%%%%%%%%%%%%%%%%%%%%%%%%%%%%%%%%%%%%%%%%%%%%%%%%%%%%%%%%%%%%%%%%%%%%%%%%%%%%%%%%%%%%%%%%%
\begin{Lemma} \label{LemmaEstimate}
Let $\Gamma$ be as in Theorem \ref{Theorem1} and let $\{t_i\}_{i=1...n} \subset \Omega$ be a set of numbers such that
$[t,t+\frac\pi 2] \cap \{t_i\} \neq \emptyset$ for all $t\in\Omega$. Assume that $|f(t_i)| \le \alpha$ for all $i$.
Then
$$\lambda_\Gamma \ge \left(1+2\alpha/\pi\right)^{-2}.$$
\end{Lemma}
Proof of Lemma \ref{LemmaEstimate}: Comparing (\ref{EqImparityOfF}) with (\ref{EqPhi}) we see that
\begin{equation}\label{EqPhiOfTPLusPi}
\phi^{-1}(t+\pi) = \phi^{-1}(t) + \pi - 2 f(t) \quad \textmd{for all } t\in \Omega.
\end{equation}
Now assume $R(s) > 0$ to be the ground state of $H_\Gamma$ and define the functions
$$x(s) := R(s) \cos \phi(s), \quad y(s) := R(s) \sin \phi(s).$$
Interpreted as Euclidean coordinates, $x$ and $y$ define a closed curve in the plane. In these coordinates the lowest
eigenvalue $\lambda_\Gamma$ of $H_\Gamma$ is
\begin{equation}\label{EqLambda}
\lambda_\Gamma = \frac{\int_\Omega \left(R'(s)^2 + \phi'(s)^2 R(s)^2 \right)\di s}{\int_\Omega R(s)^2\di s} =
\frac{\int_\Omega \left(x'(s)^2 + y'(s)^2 \right)\di s}{\int_\Omega \left(x(s)^2 + y(s)^2 \right)\di s}.
\end{equation}
We now define the orthogonal projections of the curve $(x(s),y(s))$ onto straight lines through the origin:
\begin{equation} \label{EqH}
h_\beta(s) := \begin{pmatrix} \sin \beta \\ -\cos \beta \end{pmatrix} \cdot \begin{pmatrix} x(s) \\ y(s)
\end{pmatrix} = x(s) \sin \beta - y(s) \cos \beta
\end{equation} %
We note that
\begin{equation*} %\label{EqZerosOfH1}
h_\beta(\phi^{-1}(\beta)) = 0
\end{equation*}
and, by (\ref{EqPhiOfTPLusPi}),
\begin{equation*} %\label{EqZerosOfH2}
h_\beta(\phi^{-1}(\beta) + \pi - 2f(\beta)) = h_\beta(\phi^{-1}(\beta+\pi)) = 0.
\end{equation*}
This means that the quantity
\begin{equation} \label{EqI}
I(\beta) := \frac{\int_\Omega h'_\beta(s)^2 \di s}{\int_\Omega h_\beta(s)^2 \di s},
\end{equation}%
which is the Rayleigh-Ritz quotient for the Laplacian on $\Omega$ with Dirichlet conditions at $\phi^{-1}(\beta)$ and
$\phi^{-1}(\beta) + \pi - 2f(\beta)$, can be estimated from below by
\begin{equation}\label{EqBoundOnI}
I(\beta) \ge \left(1+\frac{2 |f(\beta)|}{\pi}\right)^{-2}.
\end{equation}
Now we consider two cases: First, assume that there is no $\beta_0$ for which $I(\beta_0) = \left(1+2
\alpha/\pi\right)^{-2}$. It is clear that $I(\beta) = 1$ if $\beta$ is a zero of $f$ and we know that such a zero
exists. By continuity of $I(\beta)$ in $\beta$ we conclude that in this case $I(\beta) > \left(1+2
\alpha/\pi\right)^{-2}$ for all $\beta\in\Omega$. Choosing first $\beta = \pi/2$ and then $\beta = 0$ yields
\begin{equation*}
\frac{\int_\Omega x'^2 \di s}{\int_\Omega x^2  \di s} \ge \left(1+2 \alpha/\pi\right)^{-2} \quad \textmd{and} \quad
\frac{\int_\Omega y'^2 \di s}{ \int_\Omega y^2  \di s} \ge \left(1+2 \alpha/\pi\right)^{-2},
\end{equation*}
such that $\lambda_\Gamma \ge \left(1+2 \alpha/\pi\right)^{-2}$ by (\ref{EqLambda}).\\
In the second case there is a $\beta_0$ with $I(\beta_0) = \left(1+2 \alpha/\pi\right)^{-2}$ and by rotational symmetry
of the problem we can assume that $\beta_0 = 0$, i.e.
\begin{equation} \label{EqY}
I(0) = \frac{\int_\Omega {y'}^2 \di s}{\int_\Omega y^2 \di s} = \left(1+2 \alpha/\pi\right)^{-2}.
\end{equation}
Now put (\ref{EqH}) and (\ref{EqI}) into (\ref{EqBoundOnI}) and set $\beta = t_i$ to get
$$\int_\Omega \left(-x' \sin
t_i + y' \cos t_i \right)^2 \di s \ge \left(1+2\alpha /\pi\right)^{-2} \int_\Omega \left(-x \sin t_i + y \cos t_i
\right)^2 \di s.$$
Using (\ref{EqY}) this becomes
\begin{equation} \label{EqX}
\int_\Omega {x'}^2 \di s \ge \left(1+2\alpha /\pi\right)^{-2} \left( \int_\Omega x^2 \di s + \frac{2}{\tan
t_i}\int_\Omega (x'y'-xy) \di s\right)
\end{equation}
Because of the conditions on the distribution of the $t_i$ in $\Omega$ we can chose $i$ such that the second summand in
the bracket on the right side of (\ref{EqX}) is positive. Thus,
\begin{equation} \label{EqX2}
\int_\Omega {x'}^2 \di s \ge \left(1+2\alpha /\pi\right)^{-2}  \int_\Omega x^2 \di s.
\end{equation}
Lemma \ref{LemmaEstimate} now follows from the combination of (\ref{EqY}) with (\ref{EqX2}).
\begin{flushright} $\blacksquare$ \end{flushright}
%%%%%%%%%%%%%%%%%%%%%%%%%%%%%%%%%%%%%%%%%%%%%%%%%%%%%%%%%%%%%%%%%%%%%%%%%%%%%%%%%%%%%%%%%%%%%%%%%%%%%%%%%%%%%%%%%%%%%%%
Proof of Theorem \ref{Theorem2}: Let $\{t_i\}_{i=1...n} \subset \Omega$ be the set of critical angles of $\Gamma$. Then
by the assumption of Theorem \ref{Theorem2} this set also meets the conditions of Lemma \ref{LemmaEstimate}. Being
critical angles, the $t_i$'s satisfy $f(t_i) = 0$. Thus $\alpha$ in Lemma \ref{LemmaEstimate} can be chosen to be zero
and Theorem \ref{Theorem2} follows.
\begin{flushright} $\blacksquare$ \end{flushright}
%%%%%%%%%%%%%%%%%%%%%%%%%%%%%%%%%%%%%%%%%%%%%%%%%%%%%%%%%%%%%%%%%%%%%%%%%%%%%%%%%%%%%%%%%%%%%%%%%%%%%%%%%%%%%%%%%%%%%%%
Proof of Theorem \ref{Theorem1}: To prove Theorem \ref{Theorem1} we will derive estimates on the function $f(t)$ as
defined in (\ref{EqPhi}) and then apply Lemma \ref{LemmaEstimate}. It is obvious that we only have to consider curves
that are not covered by Theorem \ref{Theorem2}. This means that our $\Gamma$ has an interval larger than $\frac\pi 2$
without critical angles. Recall that the critical angles of $\Gamma$ are just the zeroes of $f$. We will thus assume,
without loosing generality, that $t_0$ (with $0 < t_0 < \frac\pi2$) and $\pi$ are zeroes of $f$ with a change of sign
and that $f(t) > 0$ for $t \in (t_0,\pi)$. We define $\Omega_0 := [t_0,\pi]$. Let $\Omega_+$ and $\Omega_-$ be the sets
of all points $t \in [0,t_0)$ where $f(t)$ is positive or negative, respectively. Let us now collect some
information on $f$:\\
First, we show that
\begin{equation}\label{EqBoundOnFPrime}
\int_\Omega |f'(t)| \di t \le 2\pi.
\end{equation}
To do so, we note that $(\phi^{-1})' > 0$ because we assumed $\phi' > 0$ earlier. By (\ref{EqPhi}) this means
\begin{equation*}
f'(t) + g'(t) > -1 \quad \textmd{for all} \, t\in \Omega.
\end{equation*}
But, applying (\ref{EqImparityOfF}) to this inequality, we also get
\begin{equation*}
-f'(t) + g'(t) > -1 \quad \textmd{for all} \, t\in \Omega.
\end{equation*}
Putting together the last two inequalities, we get
\begin{equation*}
|f'(t)| < 1 - g'(t) \quad \textmd{for all} \, t\in \Omega.
\end{equation*}
Integrating over $\Omega$ and keeping in mind the periodicity of $g$ yields (\ref{EqBoundOnFPrime}).\\
Second, we note that for any $\Delta, t_0\in \Omega$
\begin{eqnarray}
0 &=& \int_\Omega f(t) \sin (t + \Delta) \di t \nonumber\\
&=& \int_{t_0}^{t_0+\pi} f(t) \sin (t + \Delta) \di t +  \int_{t_0+\pi}^{t_0+2\pi} f(t) \sin (t + \Delta) \di t \nonumber\\
&=& 2  \int_{t_0}^{t_0+\pi} f(t) \sin (t + \Delta) \di t. \label{EqIntegralSinF}
\end{eqnarray}
Third, let us assume that there is an interval $[t_1,t_1+\frac\pi 2] \subset \Omega_0$ with $f(t) > \alpha$ on
$[t_1,t_1+\frac\pi 2]$ for some $\alpha \in \mathbb R$. Then
\begin{equation}\label{EqFOnOmegaPlusMinus}
\int_{\Omega_+} f(t) \di t \ge \alpha \quad\textmd{and} \quad -\int_{\Omega_-} f(t) \di t \ge \alpha.
\end{equation}
This can be seen with the help of (\ref{EqIntegralSinF}) via
\begin{eqnarray*}
0 &=& \int_0^\pi f(t) \sin(t-t_0) \di t\\
&=& \int_{\Omega_+} f(t) \sin(t-t_0) \di t +  \int_{\Omega_-} f(t) \sin(t-t_0) \di t +  \int_{\Omega_0} f(t)
\sin(t-t_0) \di t\\
&\ge& -\int_{\Omega_+} f(t) \di t + \alpha \int_{t_1}^{t_1+\pi/2} \sin(t-t_0) \di t\\
&\ge& -\int_{\Omega_+} f(t) \di t + \alpha \int_{0}^{\pi/2} \sin(t) \di t\\
&=& -\int_{\Omega_+} f(t) \di t + \alpha.
\end{eqnarray*}
The corresponding inequality for $\Omega_-$ is proven analogously, exploiting once again (\ref{EqImparityOfF}).\\
Because $f$ vanishes at the edges of $\Omega_0$ and because $\max\limits_{t\in \Omega_0} f(t) > \alpha$, it is clear
that
\begin{equation*}
\int_{\Omega_0} |f'(t)| \di t > 2 \alpha.
\end{equation*}
From (\ref{EqFOnOmegaPlusMinus}) we conclude that the inequalities
\begin{equation*}
\max\limits_{t\in \Omega_+} f(t) \ge \frac{\alpha}{|\Omega_+|} \quad \textmd{and}\quad \max\limits_{t\in \Omega_-}
|f(t)| \ge \frac{\alpha}{|\Omega_-|}
\end{equation*}
hold. Therefore
\begin{eqnarray}
\int_\Omega |f'(t)| \di t&=& 2 \left(\int_{\Omega_0} |f'(t)| \di t + \int_{\Omega_+} |f'(t)| \di t + \int_{\Omega_-}
|f'(t)| \di t\right)\nonumber\\
&\ge& 2\left( 2\alpha + 2 \frac\alpha {|\Omega_+|} + 2\frac \alpha{|\Omega_-|}\right)\nonumber\\
&>& 4 \alpha \left(1 + \frac 8\pi\right). \label{EqIntFPrime}
\end{eqnarray}
In the last step we have used $|\Omega_+|+|\Omega_-| < \frac\pi 2$. Comparing (\ref{EqIntFPrime}) with
(\ref{EqBoundOnFPrime}) shows that
\begin{equation*}
\alpha < \frac{\pi}{2(1+8/\pi)}.
\end{equation*}
We conclude, that for any curve $\Gamma$ we can can find a sequence $\{t_i\}$ that meets the conditions of Lemma
\ref{LemmaEstimate} for some $\alpha < \frac{\pi}{2(1+8/\pi)}$, proving Theorem \ref{Theorem1}.
\begin{flushright} $\blacksquare$ \end{flushright}

%%%%%%%%%%%%%%%%%%%%%%%%%%%%%%%%%%%%%%%%%%%%%%%%%%%%%%%%%%%%%%%%%%%%%%%%%%%%%%%%%%%%%%%%%%%%%%%%%%%%%%%%%%%%%%%%%%%%%
\section*{Acknowledgments}
%%%%%%%%%%%%%%%%%%%%%%%%%%%%%%%%%%%%%%%%%%%%%%%%%%%%%%%%%%%%%%%%%%%%%%%%%%%%%%%%%%%%%%%%%%%%%%%%%%%%%%%%%%%%%%%%%%%%%
It is a pleasure for me to express my gratitude to Professor Rafael D. Benguria for many helpful discussions and for
proofreading the manuscript.

\end{document}